\documentclass[journal]{IEEEtran}
\usepackage[backend=biber,
natbib=true,
style=ieee,
citestyle=numeric-comp,
sorting=none,
doi=false,
isbn=false,
url=true,
mincitenames=1,	% used for setting min number of author names in text area
maxcitenames=1,
urldate=long,	%year(2016), short(08/30/2016), long(Aug. 30, 2016), terse(08/30/2016), comp(Aug. 30, 2016), iso8601(2016-08-30)
]{biblatex}	% best for bibliography
\renewbibmacro{in:}{}
\usepackage{xpatch}	% to avoid rewriting the driver in biblatex: http://tex.stackexchange.com/questions/328128/strange-behaviour-in-biblatex-online-without-year
\xpatchbibdriver{online}
{\printtext[parens]{\usebibmacro{date}}}
{\iffieldundef{year}{}{\printtext[parens]{\usebibmacro{date}}}}
{}{}
\DefineBibliographyStrings{english}{url = [Online]\adddot\addspace Available at,}
\DefineBibliographyStrings{english}{urlseen = {Last accessed:},}
\DeclareFieldFormat{urldate}{\bibstring{urlseen}\space#1}
\DeclareFieldFormat{url}{[Online]\adddot\addspace Available at: \url{#1}\adddot}
\usepackage{url}
\usepackage{pgfplots, pgfplotstable}% loads graphicx, and xcolor
\usepackage{subcaption}
\usepackage{float,dblfloatfix} % fix for bottom-placement of figure
\usepackage{mathtools, amssymb, amsthm, nccmath}	% better math, symbols
\usepackage{booktabs}	% better tables
\usepackage{array, tabularx, tabulary, multirow, tabularht, dcolumn, longtable, siunitx, colortbl, tabu}	% table: http://tex.stackexchange.com/questions/12672/which-tabular-packages-do-which-tasks-and-which-packages-conflict
\usepackage{adjustbox}	% table scale
\usepackage{caption}	% centering caption
\usepackage{t1enc}
\usepackage{ dsfont }	%  http://www.biwako.shiga-u.ac.jp/sensei/kumazawa/tex/dsfont.html
\usepackage{soul}
\usepackage{ mathrsfs }	% math symbols
\usepackage{wasysym}	% more symbols
\usepackage[algo2e,algoruled,lined,linesnumbered]{algorithm2e}	% writting algo
\usepackage[capitalize]{cleveref}	% referencing, must be added after hyperref, [capitalise], [noabbrev] % ieee format may cause problem
\usepackage{xspace}% for macro spacing
\usepackage{balance}	% balancing last page
\usepackage[nopostdot,acronym,toc,section=chapter,nomain,automake]{glossaries}
\usepackage{cancel}

\usetikzlibrary{positioning,decorations.text,patterns}

\definecolor{Nblack}{rgb}{0,0,0}

\definecolor{color1}{HTML}{006837}
\definecolor{color2}{HTML}{31a354}
\definecolor{color3}{HTML}{78c679}
\definecolor{color4}{HTML}{c2e699}
\definecolor{color5}{HTML}{ffffcc}

\definecolor{color11}{HTML}{cc4c02}

\definecolor{color16}{HTML}{e66101}
\definecolor{color17}{HTML}{fdb863}
\definecolor{color18}{HTML}{b2abd2}
\definecolor{color19}{HTML}{5e3c99}

\definecolor{color21}{HTML}{a65628}
\definecolor{color22}{HTML}{ff7f00}
\definecolor{color23}{HTML}{984ea3}
\definecolor{color24}{HTML}{4daf4a}
\definecolor{color25}{HTML}{377eb8}
\definecolor{color26}{HTML}{666666}
\definecolor{color27}{HTML}{a6761d}
\definecolor{color28}{HTML}{e6ab02}
\definecolor{color29}{HTML}{66a61e}
\definecolor{color30}{HTML}{d95f02}
\definecolor{color31}{HTML}{1b9e77}
\definecolor{color32}{HTML}{ffd92f}
\definecolor{color33}{HTML}{7570b3}

\definecolor{color41}{HTML}{ffeda0}
\definecolor{color42}{HTML}{7fcdbb}
\definecolor{color43}{HTML}{2c7fb8}

\definecolor{color51}{HTML}{ffffbf}
\definecolor{color52}{HTML}{fee090}
\definecolor{color53}{HTML}{fc8d59}
\definecolor{color54}{HTML}{d73027}
\definecolor{color55}{HTML}{4575b4}
\definecolor{color56}{HTML}{91bfdb}
\definecolor{color57}{HTML}{e0f3f8}

\definecolor{color72}{HTML}{b59848}

\definecolor{cd41}{HTML}{e66101}
\definecolor{cd42}{HTML}{2b6d5b}
\definecolor{cd43}{HTML}{b2abd2}
\definecolor{cd44}{HTML}{5e3c99}
\definecolor{cd51}{HTML}{d7191c}
\definecolor{cs55}{HTML}{253494}

\definecolor{cfp5o1}{HTML}{fef0d9}
\definecolor{cfp5o2}{HTML}{fdcc8a}
\definecolor{cfp5o3}{HTML}{fc8d59}
\definecolor{cfp5o4}{HTML}{e34a33}
\definecolor{cfp5o5}{HTML}{b30000}

\definecolor{cft5o1}{HTML}{e41a1c}
\definecolor{cft5o2}{HTML}{377eb8}
\definecolor{cft5o3}{HTML}{4daf4a}
\definecolor{cft5o4}{HTML}{984ea3}
\definecolor{cft5o5}{HTML}{ff7f00}

\definecolor{cfp6o1}{HTML}{8c510a}
\definecolor{cfp6o2}{HTML}{d8b365}
\definecolor{cfp6o3}{HTML}{f6e8c3}
\definecolor{cfp6o4}{HTML}{c7eae5}
\definecolor{cfp6o5}{HTML}{5ab4ac}
\definecolor{cfp6o6}{HTML}{01665e}

\newcommand{\cmmnt}[1]{\iffalse #1 \fi \ignorespaces}

\makeatletter
\newcommand\resetstackedplots[1]{% argument is comma separated list of x-values
	\pgfplots@stacked@isfirstplottrue
	\pgfplotstableread[col sep=comma,row sep=crcr]{#1\\}\tmpTab
	\pgfplotstabletranspose{\tmpTab}{\tmpTab}
	\addplot [forget plot,draw=none] table[x=0,y expr=0]{\tmpTab};
	\pgfplotstableclear{\tmpTab}
}
\makeatother

\makeatletter
\newcommand{\removelatexerror}{\let\@latex@error\@gobble}
\makeatother

\makeatletter
\newcommand*{\glsplainhyperlink}[2]{%
  \colorlet{currenttext}{.}% store current text color
  \colorlet{currentlink}{\@linkcolor}% store current link color
  \hypersetup{linkcolor=currenttext}% set link color
  \hyperlink{#1}{#2}%
  \hypersetup{linkcolor=currentlink}% reset to default
}
\let\@glslink\glsplainhyperlink
\makeatother

\newcommand{\gbps}{Gbps\xspace}

\newglossary*{acronyms}{List of Acronyms}
\newglossary*{abbreviation}{List of Abbreviations}
\newglossary*{symbol}{List of Symbols}

\newglossaryentry{b5g}
{
    type=acronyms,
    name={B5G},
    first={beyond 5G (B5G)},
    description={Beyond 5G}
}
\newglossaryentry{ilp}
{
    type=acronyms,
    name={ILP},
    first={integer linear programming (ILP)},
    description={Integer linear programming}
}
\newglossaryentry{rsa}
{
    type=acronyms,
    name={RSA},
    first={routing and spectrum allocation (RSA)},
    description={Routing and spectrum allocation}
}
\newglossaryentry{nkn}
{
    type=acronyms,
    name={NKN},
    first={National Knowledge Network (NKN)},
    description={National Knowledge Network}
}
\newglossaryentry{atd}
{
    type=acronyms,
    name={ATD},
    first={average traffic demand (ATD)},
    description={Average traffic demand}
}

\makeglossaries

\cmmnt{
\author{Pramit Biswas,%~\IEEEmembership{Student Member,~IEEE,}
~Md Shahbaz Akhtar,%~\IEEEmembership{Member,~IEEE,}
~Aneek Adhya,
Sudhan Majhi, %~\IEEEmembership{Member,~IEEE,}
and~Sriparna Saha%~\IEEEmembership{Member,~IEEE,}% <-this % stops a space

\thanks{P. Biswas, Md Shahbaz Akhtar and Sudhan Majhi are with the Department of Electrical Engineering, Indian Institute of Technology Patna, India.}% <-this % stops a space
\thanks{A. Adhya is with the G.S. Sanyal School of Telecommunication, Indian Institute of Technology Kharagpur, India.}% <-this % stops a space
\thanks{S. Saha is with the Department of Computer Science and Engineering, Indian Institute of Technology Patna, India.}% <-this % stops a space
\thanks{Manuscript received Month XX, YEAR; revised Month XX, YEAR.}}
}

\cmmnt{
\author[1]{Pramit Biswas}
\author[1]{Md Shahbaz Akhtar}
\author[2,*]{Aneek Adhya}
\author[3]{Sriparna Saha}
\author[1]{Sudhan Majhi}

\affil[1]{Department of Electrical Engineering, Indian Institute of Technology Patna, India}
\affil[2]{G.S. Sanyal School of Telecommunication, Indian Institute of Technology Kharagpur, India}
\affil[3]{Department of Computer Science and Engineering, Indian Institute of Technology Patna, India}

\affil[*]{Corresponding author: aneek@gssst.iitkgp.ac.in}
}

\begin{document}
\title{Q-Learning Based Energy-Efficient Network Planning in IP-over-EON}

\author{Pramit Biswas,%~\IEEEmembership{Student Member,~IEEE,}
~Md Shahbaz Akhtar,%~\IEEEmembership{Member,~IEEE,}
~Aneek Adhya,
Sriparna Saha, %~\IEEEmembership{Member,~IEEE,}
and~Sudhan Majhi%~\IEEEmembership{Member,~IEEE,}% <-this % stops a space

%\thanks{P. Biswas, S. Akhtar and S. Majhi are with the Department of Electrical Engineering, Indian Institute of Technology Patna, India.}% <-this % stops a space
%\thanks{A. Adhya is with the G.S. Sanyal School of Telecommunication, Indian Institute of Technology Kharagpur, India.}% <-this % stops a space
%\thanks{S. Saha is with the Department of Computer Science and Engineering, Indian Institute of Technology Patna, India.}% <-this % stops a space
%\thanks{Manuscript received Month DD, YEAR; revised Month DD, YEAR.}
}

\markboth{arXiv}%
{arXiv}

\maketitle

\begin{abstract}
% Max 100 words.
% In this study, we propose a heuristic, referred to as Q-learning and AG based energy-efficient network planning for IP-over-EON (QAG-ENP-IoE) to be used for large network planning, since MILP based optimization model may not provide solution due to large computational complexity. 

During network planning phase, optimal network planning implemented through efficient resource allocation and static traffic demand provisioning in IP-over-elastic optical network (IP-over-EON) is significantly challenging compared with the fixed-grid wavelength division multiplexing (WDM) network due to increased flexibility in IP-over-EON. Mathematical optimization models used for this purpose may not provide solution for large networks due to large computational complexity. In this regard, a greedy heuristic may be used that intuitively selects traffic elements in sequence from static traffic demand matrix and attempts to find the best solution. However, in general, such greedy heuristics offer suboptimal solutions, since appropriate traffic sequence offering the optimal performance is rarely selected. In this regard, we propose a reinforcement learning
technique (in particular a Q-learning method), combined with an auxiliary graph (AG)-based
energy efficient greedy method to be used for large network planning. The Q-learning method is used to decide the suitable sequence of traffic allocation such that the overall power consumption in the network reduces. In the proposed heuristic, each traffic from the given static traffic demand matrix is successively selected using Q-learning technique and provisioned using the AG-based greedy method.
\end{abstract}


\IEEEpeerreviewmaketitle

\section{Introduction}
% https://lilianweng.github.io/lil-log/2018/02/19/a-long-peek-into-reinforcement-learning.html

%\textcolor{red}{It is anticipated that nearly two-thirds of the global population will have Internet access by 2023 \cite{CISCOAR2018to23}}. 

\arrayrulecolor{black}

With the increase in adoption of data-intensive services, such as ultra-high-definition video streaming, cloud gaming and virtual and augmented reality video streaming, global IP traffic is anticipated to increase 3-fold from 2017 to 2022~\cite{CISCOVNI2017to22}. With the increase in communication network traffic, global energy consumption in the network is expected to grow to a staggering level of 21\% of the global electricity consumption by 2030~\cite{andrae2015global}. Achieving energy-efficiency in communication networks is not practicable without paying due attention to the energy efficiency in optical backbone networks. In view of this, in this paper, we focus on energy efficiency in optical backbone networks. 

%Increasing video definition: By 2023, 66 percent of connected flat-panel TV sets will be 4K 27\% CAGR

We aim to minimize the power consumption (PC) in optical backbone network during network planning phase through efficient resource allocation and traffic provisioning taking into consideration the expected (static) traffic demands of node pairs. Elastic optical network (EON) architecture enabling the flexible grid (spectrum) allocation and use of flexible transponders [i.e., sliceable bandwidth variable transponders (SBVTs)], orthogonal sub-carriers and adaptive modulation schemes is considered to be prospective next-generation optical backbone network architecture \cite{Gerstel2012Elastic,jinno2011ip}. EON allows coexisting of multiple lightpaths with different capacities, spectrum, maximum transparent reaches (MTRs) and related PCs at SBVT in each unidirectional fiber\footnote{In this paper, unless stated otherwise, a fiber represents a bidirectional fiber with two unidirectional fibers in the opposite directions.}. In IP-over-EON architecture, IP layer is integrated with EON, and access network traffic from multiple sources are groomed through electrical layer traffic grooming in IP-core routers for onward transmission through EON~\cite{gkamas2015energy}. Energy efficiency in the network can be improved by exploiting the flexibility and reconfigurability of IP-over-EON. In the network planning phase, resource and traffic provisioning are implemented employing static traffic demand matrix, which is typically obtained from the long-term average traffic demands of node pairs or a predetermined percentile of the peak traffic demands between the nodes. 

%\textcolor{red}{EON is considered to be superior compared to the traditional fixed grid wavelength division multiplexing (WDM) optical network in view of improving flexibility, reconfigurability, spectrum efficiency, PC and scalability~\cite{jinno2009spectrum,idzikowski2016survey}.} 

%enables additional flexibility in traffic demand provisioning, i.e., multiple approaches to \textcolor{red}{allocate resources for traffic demands.}

%Since traffic profile may vary with time, in the operational phase dynamic traffic between nodes needs to be accommodated. 

%In this paper, we focus on IP-over-EON planning using the static traffic demand.
%(fixed broad band and cellular data speed may double and triple respectively by 2023 from 2018)

%, thus routing and spectrum allocation (RSA) problem is essentially become a planning problem

%Exploiting the flexibility of EON, in IP-over-EON, traffic from access network may also be groomed efficiently in IP-core routers through electrical layer traffic grooming to realize more power efficiency~\cite{gkamas2015energy}

%\textcolor{red}{\subsection{Motivation}}
In IP-over-EON, intelligent traffic provisioning may be considered as one of the approaches to improve energy efficiency in the network. The complexity to obtain the optimal network planning increases many-fold for IP-over-EON compared with the fixed-grid wavelength division multiplexing (WDM) networks due to increased flexibility in IP-over-EON. To provision traffic demands (elements) from the static traffic demand matrix, a mathematical optimization model may be used for optimal network planning~\cite{krishnaswamy2001design,zhang2015energy,biswas2020energy}. However, the optimization model may not provide solution for large networks in presence of all related constraints due to large computational complexity. Thus, a greedy heuristic may be used that intuitively selects one traffic demand at a time (i.e., at each step) and routes the traffic~\cite{biswas2019energy,zhang2015energy}. The process is repeated for all traffic elements in the traffic demand matrix with the traffic elements selected in a sequence, as the heuristic attempts to obtain the best solution. However, the greedy heuristic has only one chance to select and route a traffic, and never reviews the decision taken at an earlier step. It may offer the best performance for specific traffic sequence(s) only, and any other traffic sequence will yield suboptimal solution. As for example, Zhao \textit{et al.} show that different sequences of traffic demands, such as traffic demands with decreasing order of data-rate requirement or decreasing order of the number of fiber links along the shortest path between the source and destination (SD) nodes may change the network performance in terms of the maximum allocated sub-carrier indices on a fiber~\cite{zhao2015nonlinear}. Thus, in the planning phase, there lies opportunity for optimization of network performance by deciding the best sequence of static traffic demands to be selected. If in a given network with $N$ number of nodes, traffic demands exist among all nodes in the static traffic demand matrix, i.e., $N(N\text{-}1)$ number of traffic exist (excluding the diagonal elements), there can be $[N(N\text{-}1)]!$ numbers of possible sequences following which the traffic demands can be provisioned. Therefore, the search space is significantly large for moderate to large value of $N$. An analogy can be made between this problem and finding the shortest Hamiltonian path problem, which is considered to be NP-hard. A Hamiltonian path ensues that travelling along the path, all nodes in the graph are visited only once, similar to provisioning all traffic elements only once. However, one important difference between the two problems is that for the shortest Hamiltonian path, the distance between any two nodes (i.e., the related cost in traversing between the nodes) is fixed irrespective of the order (i.e., sequence) of traversal of the nodes, whereas in case of network planning problem, the cost value is not fixed, and it depends on the existing network condition, i.e., the existing lightpath status, resource availability etc.

Reinforcement learning (RL) is a category of machine learning (ML) technique that may be used to make a sequence of decisions. Watkins first propose Q-learning algorithm \cite{watkins1989learning}, which is an RL technique, and the convergence of the algorithm is proved in~\cite{watkins1992q}. Gambardella \textit{et al.} propose Q-learning based algorithms in a weighted complete graph to find the shortest Hamiltonian tours, analogous to determining the sequence of static traffic demands to be selected for optimal provisioning~\cite{gambardella1995ant}. For efficient network planning, we explore a Q-learning based technique, which is a reward based trial and learn method and does not require any labeled data.

%We explore Q-learning based technique for optimal network planning and traffic provisioning since it is a reward based trial and learn method which does not require any labeled data.

%Reinforcement learning (RL) algorithm may be used to train ML models to make a sequence of decisions. Q-learning algorithm, that is an RL technique is first proposed by C.~Watkins in~\cite{watkins1989learning}, and the convergence of the algorithm is proved in~\cite{watkins1992q}. Gambardella \textit{et al.} propose Q-learning based algorithms to find the shortest Hamiltonian tours in a weighted complete graph~\cite{gambardella1995ant}. 

%In this paper, the possible application of machine learning (ML) techniques is explored for the above-mentioned scenario.

%Thus, in general, following the method only suboptimal solution can be obtained. 

%In general, a greedy heuristic approach is considered to provision single traffic demand, which gives optimized or near to optimized results.

%\textcolor{red}{\st{In the dynamic traffic provisioning scenario, the number of traffic to be provisioned at a given time duration is low compared to the static scenario where traffic between all nodes need to provision.}}
% crucial
% irrespective of their occurrence
% depends  on  previously  provisioned  traffic  sequence. 

%In this paper, possible application of Q-learning algorithm is explored for network planning and traffic provisioning problem in IP-over-EON.

In this paper, we explore energy-efficient network planning for IP-over-EON using static traffic demands by deciding the best sequence of traffic to be provisioned using Q-learning technique combined with a greedy heuristic. Each traffic from the given static traffic demand matrix is successively selected using Q-learning technique and provisioned using greedy heuristic. The process is repeated for the traffic demand matrix for multiple times, and the best network planning is identified. We use an auxiliary graph (AG)-based energy-efficient greedy heuristic to provision the selected traffic with the least increase in PC. For identification of traffic sequence, Q-learning based algorithm is used, where an agent predicts an action, i.e., the next traffic to be provisioned from the past experiences, and gains reward based on the PC needed for provisioning. From the received rewards, the agent tries to develop the optimal policy that helps to decide the best sequence for energy-efficient traffic provisioning. The performance of the proposed Q-learning and AG based energy-efficient network planning for IP-over-EON heuristic, referred to as QAG-ENP-IoE, is assessed with realistic network setting. Moreover, we also use virtualized elastic regenerators (VERs) to enhance flexibility,  connectivity,  and improving energy efficiency in the network.

%In this paper, the sequence prediction is explored using genetic algorithm and reinforcement learning along with existing greedy heuristic for energy-efficient static IP-over-EON network planing.

%\subsection{Paper Organization}
In \cref{c4:rw}, related literature are presented. Next, in \cref{c4:s:na}, a brief description of IP-over-EON architecture and PC model for network equipment are presented. In \cref{c4:QAGENP}, the proposed heuristic is described, while in \cref{c4:papa} the performance of the heuristic is studied. Finally, \cref{c4:conclusion} concludes the paper.

\section{Related Work}\label{c4:rw}
%basic static algo
%energy efficient algo
%genetic algo/ants colony etc static

Klinkowski \textit{et al.} propose an \gls{ilp} based optimization model to solve \gls{rsa} problem with the objective to minimize the use of spectrum resources, i.e., frequency slots (FSs) for optimal network planning in EON using static (offline) traffic demands~\cite{klinkowski2011routing}. Since the optimization model is NP-hard, a greedy heuristic is proposed for large problem size where traffic demands are allocated in the decreasing order of requested FSs. Zhang \textit{et al.} propose a heuristic to maximize optical layer traffic grooming in IP-over-EON  facilitating simultaneous generation/termination of multiple lightpaths of different capacities, FSs and MTRs by a single SBVT, so as to minimize the overall PC \cite{zhang2015energy}. The proposed heuristic provisions traffic in descending order of the requested bandwidth. The authors in ~\cite{biswas2019energy} propose an AG-based heuristic for energy efficient network planning in IP-over-EON, where the sequence of traffic to be processed is determined following the descending order of traffic bandwidth. Furthermore, a pruning strategy is used to further reduce the PC wherever possible. Ramaswami \textit{et al.} propose an mixed ILP (MILP) model for network planning in fixed-grid WDM networks~\cite{ramaswami1996design}. The MILP model is further decomposed into virtual topology design and traffic routing sub-problems, and the solution from the first sub-problem is provided as input to the second sub-problem so as to improve computational tractability for large networks. Zhao \textit{et al.} explore an ILP formulation to solve nonlinear impairment-aware RSA problem for EON with the objective to reduce the maximum index number of FSs to be used on a fiber \cite{zhao2015nonlinear}. In this regard, simulated annealing based heuristic is also explored for large problems following three different traffic demand ordering policies, viz., decreasing order of traffic demands, decreasing order of the number of links along the shortest paths, and decreasing order of the product of shortest path length and traffic demand \cite{zhao2015nonlinear}. In~\cite{biswas2016auxiliary}, with reference to dynamic traffic provisioning in IP-over-EON, different job scheduling strategies (i.e., identifying sequence of traffic demands) are adopted based on the bandwidth and holding time criteria, and their impact on the overall energy efficiency is studied.

%  We develop a framework that leverages the probabilistic outputs of a ML based QoT estimator to define the reach constraints in an Integer Linear Programming (ILP) formulation for RSA in an elastic optical network. In this integrated procedure, the RSA problem is solved iteratively by updating the reach constraints based on the outcome of a QoT estimator, to exclude lightpaths with unacceptable QoT.
 
%  QoT estimator which iteratively used by an ILP model 

% either a feasible solution for all lightpaths is found, or a maximum number of iterations is reached

Musumeci \textit{et al.} discuss different application areas in optical networking domain, such as path computation, dynamic traffic prediction, failure management, and  quality of transmission (QoT) estimation, where ML techniques can be used~\cite{musumeci2018overview}. Salani \textit{et al.} present an ML classifier based QoT estimator to estimate parameters in transmission reach constraints of an ILP model, in order to solve the routing, modulation format and spectrum assignment problem~\cite{salani2019routing}. Following iterative approach, the method excludes lightpaths with poor QoT as estimated by the ML classifier, until either a feasible solution for all lightpaths is found, or the upper limit of iteration count is reached. In~\cite{lian2020flexible}, to achieve a fast network recovery from an IP node failure in IP-over-EON, a Q-learning based recovery algorithm is presented. Kiran \textit{et al.} propose algorithms based on Q-learning to solve path selection and wavelength selection in optical burst switch (OBS) networks with objective to minimize the burst loss probability~\cite{kiran2007reinforcement}. The algorithm is used to select path and wavelength from a set of pre-computed paths and a set of wavelengths, respectively. In case of IP node failure in 5G and \gls{b5g} IP-over-optical network, Gu \textit{et al.} use Q-learning based algorithm to re-configure the optical layer in view of service recovery, without requiring to re-route the affected traffic flows individually~\cite{gu2020flexible}. The proposed method helps in mitigating exhaustive IP forwarding and routing requirement. To the best of our knowledge, identifying the sequence of traffic for network planning in IP-over-EON using any ML technique has not been studied till date.

\section{Network Architecture and PC model for Network Equipment}\label{c4:s:na}

\begin{figure}[t]%[b!]%[b!]%[h]%htbp!
\centering
\includegraphics[width=0.4\textwidth]{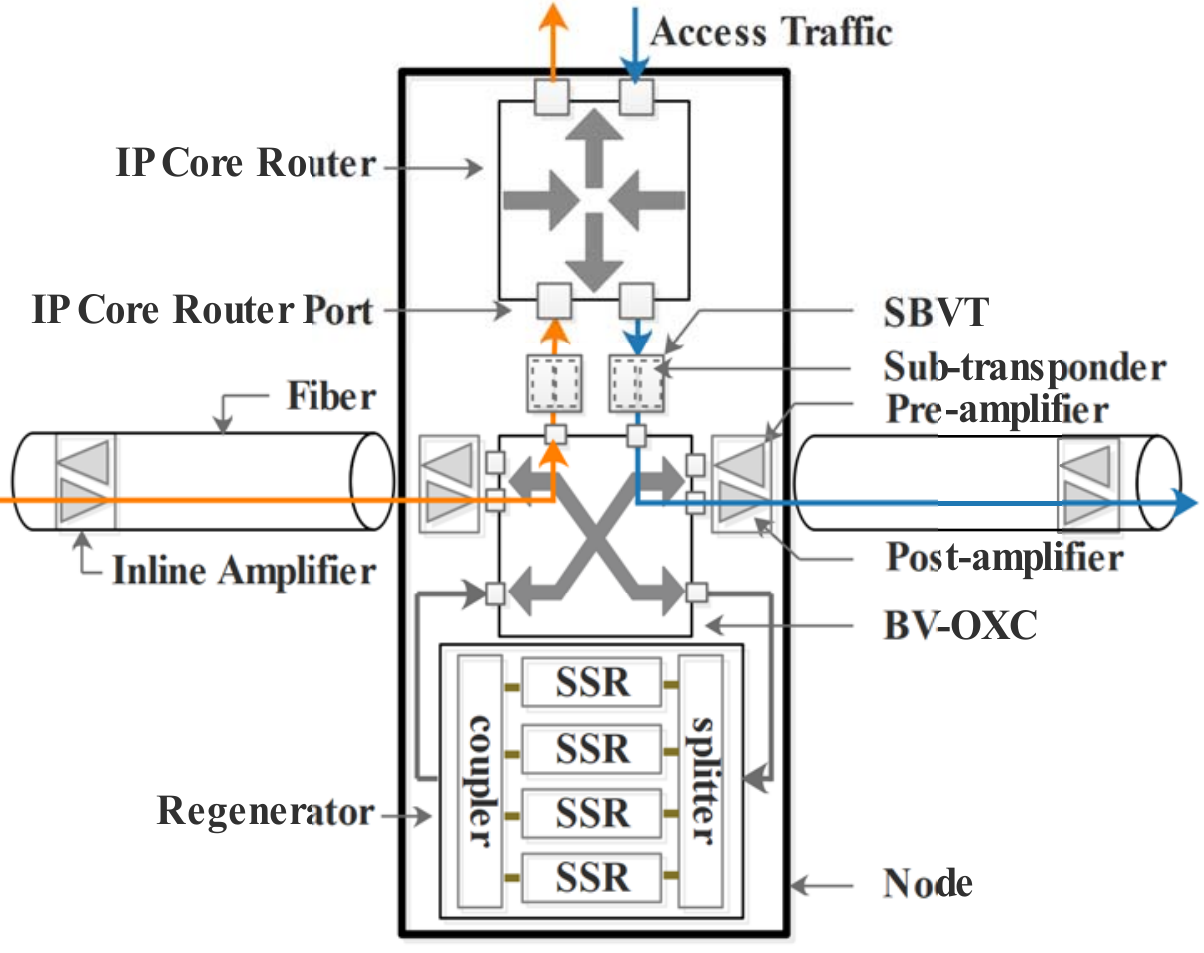}
\captionsetup{justification=centering,font={small}}%,margin=1cm,font={footnotesize}, scriptsize, footnotesize
\caption{An IP-over-EON node architecture.}
\label{c4:ipoena}
\end{figure}

The IP-over-EON architecture has a partial mesh topology where nodes are connected with fibers, and fibers have inline optical amplifiers installed with a fixed span (regular interval) of distance $L$. As shown in \cref{c4:ipoena}, at each node, a post-amplifier is connected with the outbound unidirectional fiber, whereas a pre-amplifier is connected with the inbound unidirectional fiber. An optical amplifier placed at a location is composed of two unidirectional amplifiers operating at opposite directions, along with the related electronic circuit. Thus, optical amplifiers are treated as bidirectional. We consider the following PC model for an optical amplifier that each optical amplifier has a constant overhead PC of $P_{A_O}$ along with a fixed PC of $P_{A_d}$ for each unidirectional amplifier~\cite{vizcaino2012energyArticle,biswas2019energy}. As shown in \cref{c4:ipoena}, each node has one IP core router where each port has a fixed capacity of $C_R$ and a fixed PC of $P_R$~\cite{van2012power,biswas2019energy}. Each router port is connected to a SBVT that is equipped with multiple sub-transponders with each sub-transponder supporting transmission/reception of one lightpath. Each lightpath is characterized by multiple attributes, such as the capacity, the MTR, the number of data-slots and the required PC at SBVT. We consider that different lightpath transmission options are available at SBVT following \cref{c4:t:lto} where each data-slot is considered to be of 12.5~GHz \cite{biswas2019energy,biswas2019edfa,papanikolaou2015minimizing}. PC to transmit/receive a lightpath at SBVT is considered to be half of the PC of an active sub-transponder~\cite{zhang2015energy,biswas2019energy}. Each SBVT is connected to an add-drop port of a bandwidth variable optical cross-connect (BV-OXC). A BV-OXC has the same PC model as that of the OXC used in fixed-grid WDM networks. The PC in a BV-OXC is considered to be (135.$d$+150) W for 100\% add-drop facility, where $d$ is the physical degree of the node accommodating the BV-OXC. Some of the nodes may also have one or more VERs~\cite{Jinno2012Elastic,Jinno2012Demonstration} connected to BV-OXC for 4R-regeneration (re-shaping, re-timing, re-amplifying, and re-modulation) of lightpaths and merging of FSs when routes of multiple lightpaths after regeneration are the same. A VER consists of a splitter, a coupler and an array of spectrum selective regenerators (SSRs)~\cite{Jinno2012Elastic,Jinno2012Demonstration}. We consider that PC in a VER depends on the lightpaths being regenerated through SSRs, where PC for lightpath regeneration is same as the summation of the PC for termination and generation of the same type of lightpath through sub-transponders in SBVT (as shown in \cref{c4:t:lto})~\cite{biswas2019edfa}. We consider that overhead PC for each SSR is $P_{V_S}$, and the number of SSRs used in a VER depends on the number and type of lightpaths regenerated. The overhead PC for each VER is considered to be $P_{V_O}$~\cite{biswas2020energy}. The total PC in a VER is the summation of the PC for the SSRs used and the overhead PC~\cite{biswas2020energy}.

% \textcolor{red}{The PC in a VER is the summation of the PC for all lightpaths being regenerated therein, the PC for the SSRs used and the overhead PC~\cite{biswas2019edfa}}. \textcolor{blue}{PC for regeneration of a lightpath at VER is considered to be same as the summation of the PC for termination and generation of the same type of lightpath through sub-transponders in SBVT (as shown in \cref{c4:t:lto})~\cite{biswas2019edfa}.} PC for each SSR is considered to be fixed ($P_{V_S}$), while the overhead PC for each VER is considered to be $P_{V_O}$~\cite{biswas2019edfa}.

%A VER uses an array of spectrum selective regenerators (SSRs) for regeneration.

\begin{figure}
\centering
\captionof{table}{Multiple transmission options for sub-transponder of an SBVT and VER~\cite{biswas2019energy,biswas2019edfa,papanikolaou2015minimizing}}% and regenerator
\label{c4:t:lto}
\begin{adjustbox}{max width=\linewidth, max height=3cm}
\begin{tabular}{@{}cccccccc@{}}%||
\cmidrule(r){1-4}\cmidrule(l){5-8}%\toprule

\begin{tabular}[c]{@{}c@{}}Capacity\\ (Gbps)\end{tabular} & \begin{tabular}[c]{@{}c@{}}MTR\\ (km)\end{tabular} & \begin{tabular}[c]{@{}c@{}}Data\\ Slot\end{tabular} & \begin{tabular}[c]{@{}c@{}}PC\\ (W)\end{tabular} & \begin{tabular}[c]{@{}c@{}}Capacity\\ (Gbps)\end{tabular} & \begin{tabular}[c]{@{}c@{}}MTR\\ (km)\end{tabular} & \begin{tabular}[c]{@{}c@{}}Data\\ Slot\end{tabular} & \begin{tabular}[c]{@{}l@{}}PC\\ (W)\end{tabular} \\ \cmidrule(r){1-4}\cmidrule(l){5-8}%\midrule
\multirow{5}{*}{40} & 600 & 1 & 154.8 &\multirow{5}{*}{100} & 600 & 1 & 198 \\
& 1900 & 1 & 183.6 & & 1900 & 1 & 270 \\
& 2500 & 2 & 183.6 & & 2500 & 2 & 270 \\
& 3000 & 3 & 183.6 & & 3000 & 3 & 270 \\
& 4000 & 4 & 183.6 & & 3500 & 4 & 270 \\ \addlinespace%\midrule
\multirow{6}{*}{200} & 500 & 1 & 333 &\multirow{6}{*}{400} & 500 & 4 & 432 \\
& 600 & 2 & 333 & & 600 & 6 & 432 \\
& 750 & 3 & 333 & & 750 & 8 & 432 \\
& 1900 & 4 & 432 & & 1900 & 10 & 630 \\
& 2200 & 5 & 432 & & 2200 & 12 & 630 \\
& 2500 & 6 & 432 & & 2500 & 14 & 630 \\ \bottomrule
\end{tabular}
\end{adjustbox}
\end{figure}

\section{Proposed Heuristic: QAG-ENP-IoE}\label{c4:QAGENP}
\subsection{Problem Description}
In the proposed heuristic, our objective is to minimize the overall PC in an IP-over-EON with the given network topology and static traffic demands of node pairs. The outcome (solution) of the heuristic is to determine the network resources (viz., IP core router ports, SBVTs and VERs) to be provisioned at the appropriate node locations, the setting up of lightpaths and allocation of traffic demands through lightpaths (i.e., traffic routing). The PC in the network is also computed. Typically, for this purpose, mathematical optimization models are used for small networks, whereas greedy heuristics are used for large networks. In this study, Q-learning technique integrated with a greedy heuristic is explored for large networks so as to reduce the overall PC as estimated using any greedy heuristic. The greedy heuristic we use for this purpose is discussed in the following.

%for finding the possibility of further reduction

\subsection{Greedy Heuristic}

\begin{figure}[b]%[b!]%[b!]%[h]%htbp!
\centering
\includegraphics[width=0.4\textwidth]{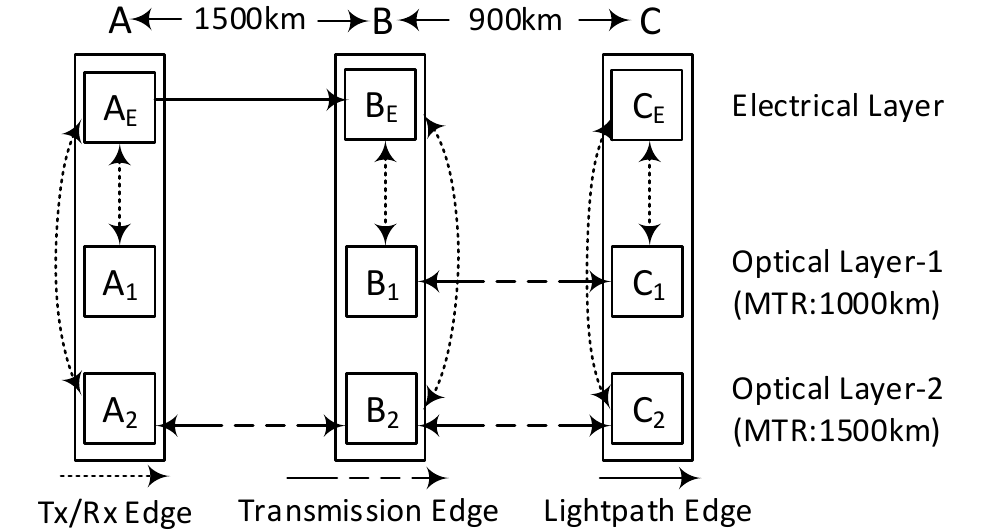}
% \captionsetup{justification=centering,font={small}}%,margin=1cm,font={footnotesize}, scriptsize, footnotesize
\caption{An example auxiliary graph~\cite{biswas2019energy,biswas2019edfa}.}
\label{c4:ag}
\end{figure}

We use modified AG-based energy-efficient network planning for IP-over-EON (mAG-ENP-IoE) heuristic (\cref{c4:a:greedy}) based on~\cite{biswas2019energy} as the greedy heuristic. Using the heuristic we provision the selected traffic demands, one at a time, in an energy-efficient manner. A brief description of the heuristic is presented in the following. 

For a given traffic demand, first, we determine the capacity of the lightpath (that may required to be set up) as the closest (equal or higher) transmission rate supported by an SBVT with reference to the traffic demand. Thereafter, with reference to the traffic demand we construct an AG (\cref{c4:ag}), as also described in~\cite{biswas2019energy}. In this regard, we consider a physical node to be composed of an electrical layer auxiliary node (AN) $A_E$, and multiple optical layer ANs, viz., $A_1$, $A_2$, etc., corresponding to different available transmission options in SBVTs. AG (\cref{c4:ag}) is constructed using different types of edges, viz., transmission edges, Tx/Rx edges and lightpath edges. Transmission edge is set up between two similar type optical layer ANs located at two different physical nodes if the required FSs for possible lightpath set up are available in the connecting fiber. The transmission edge weight is represented by the PC of optical amplifiers to be used on the fiber. Tx/Rx edge is set up from (to) an electrical layer AN to (from) an optical layer AN within the same physical node if the sub-transponder with required capacity is available at SBVT. The related edge weight is represented by the PC of IP core router and SBVT to originate (terminate) the lightpath. Lightpath edge is set up between two electrical layer ANs located at two different physical nodes if free capacity in the existing lightpath between the two physical nodes is available to provision the traffic demand. Lightpath edge weight is represented by the PC to accommodate the traffic demand to an existing lightpath. As for an example, in \cref{c4:ag}, we show three physical nodes $A$, $B$ and $C$, with distances between $A$ and $B$, and $B$ and $C$ of 1500~km and 900~km, respectively \cite{biswas2019energy,biswas2019edfa}. We consider availability of only two transmission options in the network with MTRs of 1000~km and 1500~km. We show several AG edges: Tx/Rx edges (viz., $A_E$-$A_1$, $A_E$-$A_2$ etc.), lightpath edge (viz., $A_E$-$B_E$) and transmission edges (viz., $A_2$-$B_2$, $B_2$-$C_2$ etc.)

Next, using the AG, the shortest path, i.e., the path with the minimum PC between the  electrical layer ANs of the SD nodes of the traffic demand is determined. If any path is not available, the traffic demand is split into two traffic elements between the same SD pair, such that at the least one traffic element becomes the closest transmission rate supported by an SBVT, and the algorithm starts again from Line 1 for possible provisioning of each split traffic one by one. If a path is available and the path consists of one or more lightpath edges, the traffic is groomed with the current traffic in the existing lightpath(s). On the other hand, if the path consists of one or more Tx/Rx edges, new lightpath(s) are set up with due consideration of possible utilization of VER in the path to improve energy-efficiency. Finally, the traffic demand is accommodated and the increase in PC due to provisioning of the traffic demand through the path is computed. If the traffic demand is provisioned fully, $true$ is returned; else, $false$ is returned.

%Choice of greedy algorithm can be \textcolor{red}{any} and is \textcolor{red}{fully based on problem objective}. 
%From AG-ENP-IoE of~\cite{biswas2019energy}, the initial ordering of traffic and pruning are excluded for this scenario as modification.

% \begin{figure}[h]
% \removelatexerror
% \centering
% \begin{adjustbox}{scale=1}%max width=\linewidth, max height=6cm%width=\linewidth,height=1.5cm
	\begin{algorithm2e}%[H]%hbtp!
\caption{Modified AG-based energy-efficient network planning for IP-over-EON (mAG-ENP-IoE)}\label{c4:a:greedy}

%\While{$all \ t^{sd} \neq 0$}{	\label{sA1}
	Set the capacity of lightpath that may need to be set up between SD node pair.\label{c4:sA3}\\
	Construct AG.\label{c4:sA4}\\
	Find shortest path between SD node pair.\label{c4:sA5}\\
	\eIf{no path available \label{c4:sA6}}{
		Split traffic demand into two traffic elements and start again from Line 1 to provision all split traffic.
	}{\label{c4:sA8}
		\If{Lightpath edge(s) appear in the path\label{c4:sA9}}{
			Groom traffic element in the existing lightpath(s).
		}
		\If{Tx/Rx edge(s) appear in the path\label{c4:sA12}}{
			Set up new lightpath(s).
		}\label{c4:sA14}
		Accommodate traffic.\label{c4:sA15}
	} \label{c4:sA17}
%} 
Compute PC due to traffic flow through the path.\\\label{c4:sA19}
\eIf{traffic provisioned fully}{
return $true$
}{
%reset all variables to previous state.\\
return $false$
}
\end{algorithm2e}
	%selected $\lambda^{sd}$ 
% \end{adjustbox}
% \end{figure}

\subsection{Q-Learning}

The main idea of Q-learning is to develop a policy to take certain actions based on the state of an environment or system. In Q-learning based algorithms, agents learn the policy based on rewards received by taking actions in different states, and the objective is to maximize the overall reward. Thus, to maximize the overall reward, agent focuses on long-term rewards rather than focusing only on the immediate reward. Next, we discuss Q-learning \cite{sutton2018reinforcement} in the following in brief.

There are three sets in Q-learning, viz., the set of states $\textbf{S}$ in the environment, the set of actions $\textbf{A}$ from which an agent can take an action, and the set of rewards $\textbf{R}$ which comprises the reward received by the agent for each action. At each step $t$, agent is in a representative state $s\in\textbf{S}$ in the environment where the agent takes an action $a\in\textbf{A}$ forming a state-action pair $(s, a)$. For action $a$ taken in the state $s$ at step $t$, the agent in the next step (i.e., $t+1$) reaches to state $s'\in\textbf{S}$ and receives reward $R_{t+1}\in\textbf{R}$. At step $t$ the expected overall reward ($G_t$) is the summation of all future rewards:

%As a result, the agent in the next step (i.e., $t+1$) reaches to state $s'\in\textbf{S}$ and receives reward $R_{t+1}\in\textbf{R}$ for the action taken (i.e., $a$) in the previous state (i.e., $s$) at step $t$.

\begin{equation}
    G_t = R_{t+1}+R_{t+2}+R_{t+3}+...\label{c4:e:sr}
\end{equation}

However, it may happen that in order to gain immediate high reward the agent may fail to win distant future high reward. Thus, rather than optimizing \cref{c4:e:sr}, discounted summation is optimized [\cref{c4:e:dsr}]. Here, $\gamma\in[0,1]$ represents the discount rate; having value close to $0$ shows the focus of agent is on short term rewards and value close to $1$ shows focus on long term rewards.

% \textcolor{red}{The discount rate determines the present value of future rewards: a reward received $k$ time steps in the future is worth only $\gamma^{k\text{-}1}$ times what it would be worth if it were received immediately.}

\begin{equation}
\begin{aligned}
    G_t &= R_{t+1}+\gamma R_{t+2}+\gamma^2 R_{t+3}+...\\
    &= R_{t+1}+\gamma G_{t+1}  \label{c4:e:dsr}
\end{aligned}
\end{equation}
% \textcolor{red}{= \sum^{\infty}_{k=0}\gamma^k R_{t+k+1}}

The probability of selecting an action at a particular state is determined by the policy function, i.e., following policy $\pi$ at step $t$ on state $s$, the probability of selecting action $a$ is $\pi(a|s)$. The merit of any state for policy $\pi$, i.e., the value of a state if policy $\pi$ is followed, can be determined by state-value function $v_\pi$ using \cref{c4:e:svf}, where $\mathbb{E}_\pi$ denotes the expected value of random variable when agent follows policy $\pi$ and $S_t$ denotes the state at step $t$.

%\textcolor{blue}{For each state, $s\in \textbf{S}$, $\pi$ is the probability distribution over $a\in \textbf{A(s)}$}

% \st{which is equal to the expected return from the starting state $s$ at time $t$ and following policy $\pi$ thereafter}

% random variable: https://web.stanford.edu/class/psych209/Readings/SuttonBartoIPRLBook2ndEd.pdf, p84, next line after Eq. (3.10)

\begin{equation}
\begin{aligned}
    v_\pi(s)=\mathbb{E}_{\pi} \Bigg[\sum^{\infty}_{k=0}\gamma^k R_{t+k+1}|S_t=s\Bigg]\label{c4:e:svf}%|S_t=s
\end{aligned}
\end{equation}

Similarly, merit of any action at a state for policy $\pi$, i.e., the value of an action under policy $\pi$ can be determined by action-value function $q_\pi$ using \cref{c4:e:avf}, where $A_t$ denotes the action at step $t$. This shows the quality of taking an action at a state, also known as the Q-function, and the obtained value is referred to as Q-value for the state-action pair.

\begin{equation}
\begin{aligned}
    q_\pi(s,a)=\mathbb{E}_{\pi} \Bigg[\sum^{\infty}_{k=0}\gamma^k R_{t+k+1}|S_t=s,A_t=a\Bigg]\label{c4:e:avf}%|S_t=s,A_t=a
\end{aligned}
\end{equation}

A policy $\pi$ can be said to be better than any other policy $\pi'$ if and only if $v_\pi(s)\ge v_{\pi'}(s), \forall s\in \textbf{S}$. The optimal state-value function $v_*(s)$, where $v_*(s)=\max_\pi v_\pi(s), \forall s\in\textbf{S}$ shows the highest expected reward by any policy at each state. Similarly, the optimal action-value function, or optimal Q-function $q_*(s,a)$, where $q_*(s,a)=\max_\pi q_\pi(s,a), \forall s\in\textbf{S},a\in\textbf{A}$ shows the highest expected reward by any policy for each possible state-action pair.

Hence, for any state-action pair $(s,a)$ at step $t$, the expected reward for selecting an action $a$ at state $s$ following optimal policy is the summation of expected reward for taking action $a$ at state $s$ (i.e., $R_{t+1}$) and the maximum expected discounted reward that can be achieved from any possible next state-action pair $(s',a')$.

% reward for selecting an action $a$ at state $s$ following} 

\begin{equation}
\begin{aligned}
    \text{i.e., }q_*(s,a)=\mathbb{E}\Bigg[R_{t+1}+\gamma\max_{a'}q_*(s',a')\Bigg]\label{c4:e:bme}
\end{aligned}
\end{equation}

\cref{c4:e:bme} is known as Bellman equation for $q_*$. In this study, Q-learning method is used to learn the optimal Q-value for each state-action pair in order to find the optimal policy. In Q-learning, iteratively Q-values are updated for each state-action pair using Bellman equation to converge Q-function towards the optimal Q-function ($q_*$). This approach is known as value iteration. Thus, from optimal $q_*$, optimal policy can be determined, as at any state $s$ action $a$ can be found using Q-learning algorithm that maximizes $q_*(s,a)$.

Over the time the agent goes through multiple episodes (iterations). In each episode, agent takes action at every step based on the highest Q-value of the present state (i.e., performs exploitation), and updates state and Q-value accordingly until the stopping criteria of the episode is met. Initially, Q-values for all state-action pair are maintained at zero value, and thus agent explores different states (i.e., performs exploration). To balance between exploration and exploitation, initially exploration rate ($\epsilon$) is typically set to $1$ and gradually decayed (having decay rate $\epsilon_\Delta$) with increase in episodes. The exploration rate may be determined following \cref{c4:e:eRu}, where $\epsilon_{\min}$ and $\epsilon_{\max}$ are set to $0$ and $1$ respectively. At each step, a random number $[0,1]$ is generated and compared with $\epsilon$ to select between the exploration or exploitation modes. In exploration mode, an action is selected randomly and this strategy is referred to as epsilon greedy strategy.

\begin{equation}
\begin{aligned}
    \epsilon = \epsilon_{\min}+ (\epsilon_{\max} - \epsilon_{\min}) e^{-\epsilon_{\Delta}\cdot episode} \label{c4:e:eRu}
\end{aligned}
\end{equation}

How quickly agent discards the already learned Q-value is controlled by the learning rate $\alpha\in[0,1]$. Lower the learning rate, more slowly the agent updates the Q-value. Thus, from \cref{c4:e:bme}, for state-action pair $(s,a)$ at step $t$, the new Q-value, as shown in \cref{c4:e:qu} is computed as the weighted sum of the previous Q-value and the learned value.

\begin{equation}
\begin{aligned}
    q^{new}(s,a) & = (1-\alpha)q(s,a)\\
    & + \alpha\big[R_{t+1}+\gamma \max_{a'}q(s',a')\big]\label{c4:e:qu}
\end{aligned}
\end{equation}

\subsection{QAG-ENP-IoE}

In \cref{c4:a:spq}, we present QAG-ENP-IoE heuristic where we sequentially identify a traffic using Q-learning technique and provision it using the AG-based greedy heuristic (i.e., mAG-ENP-IoE). Here, at a given episode, a state signifies the 
% \begin{figure}[!b]
% \removelatexerror
% \centering
% \begin{adjustbox}{max width=\linewidth}%max width=\linewidth, max height=6cm%width=\linewidth,height=1.5cm,scale=1
% \begin{algorithm}{!b}
	\begin{algorithm2e}%{b}%[H]%hbtp!
		\caption{Q-learning and AG-based energy-efficient network planning for IP-over-EON (QAG-ENP-IoE)}\label{c4:a:spq}
		
		Initialize all parameters. \label{c4:a:spq:l1}
		
		\For{$episode\gets0$ \KwTo $totalEpisodes$-$1$}{
            state = 0, doneFlag = False
            
            \For {$step\gets0$ \KwTo $maxStepsPerEpisode-1$}{
            
            explorationRateThreshold = random.uniform(0, 1)
            
            \eIf{ explorationRateThreshold > $\epsilon$}{
            action = action having max Q-value in current state except already taken actions.
            }{
            action = select random action except already taken actions.
            }
            Use \cref{c4:a:greedy} to provision traffic.
            
            \eIf{trafficProvisioned}{
            reward = -(increment in overall PC)\\
            \If{allTrafficProvisioned}{
            doneFlag = True\\
            reward += $R$
            }
            }{
            doneFlag = True\\
            reward = -$P$
            }
            
            update Q-value using \cref{c4:e:qu}
            
            state += 1
            
            \If{doneFlag}{
                break
            }
            }
            update $\epsilon$ using \cref{c4:e:eRu}
        }
	\end{algorithm2e}
% 	\end{algorithm}
	%selected $\lambda^{sd}$ 
% \end{adjustbox}
% \end{figure}
number of traffic that already have been provisioned, and action signifies taking attempt to provision the next traffic. First, all parameters are initialized: all Q-values are set to zero, the total number of actions and the maximum number of steps per episode are made equal to the total number of traffic to be provisioned (i.e., all non-zero elements in the traffic demand matrix), the total number of states is made equal to the total number of traffic to be provisioned added with two (representing the initial state when no traffic has been provisioned and the final state when all traffic have been provisioned), and the total number of episodes, $\alpha$, $\gamma$, $\epsilon_{\min}$, $\epsilon_{\max}$ and $\epsilon_{\Delta}$ values are provided.

%Following the above description, a Q-learning and AG-based energy-efficient heuristic (viz., QAG-ENP-IoE) is presented in \cref{c4:a:spq}

For each episode, the agent first initializes the starting state to zero and $doneFlag$ to false. The execution of the current episode is terminated (stopped) in case all traffic have been provisioned or a given traffic cannot be provisioned. If any of these two conditions is met, $doneFlag$ is set to true. Next, for each step, first, the action based on exploration or exploitation is decided. A random number [0,1] is generated and compared with $\epsilon$. In case the random number is greater than $\epsilon$, an unprovisioned  traffic is selected based on the maximum Q-value at the current state. Otherwise, an unprovisioned traffic is selected through uniform random distribution. Next, attempt is made to provision the selected traffic  using mAG-ENP-IoE (\cref{c4:a:greedy}). If the traffic is provisioned successfully, the incurred reward is the negative of the increase in PC, else reward is -$P$, where, $P$ represents a very large number used as penalty. Reward is considered to be negative as our problem to reduce PC is a minimization problem. Now, for unsuccessful traffic provisioning or in case all traffic have been provisioned, $doneFlag$ is set to true. Additionally, in case all traffic can be provisioned, a high reward $R$ is added to the existing reward. Next, corresponding Q-value is updated following \cref{c4:e:qu} and state value is increased by one. If $doneFlag$ is $true$ then the current episode ends. After completion of each episode, $\epsilon$ value is updated following \cref{c4:e:eRu}.

\section{Performance Analysis of Proposed Heuristic}\label{c4:papa}

\begin{figure}[h!]%[b!]%[b!]%[h]%htbp!
% \centering
\includegraphics[width=.9\linewidth]{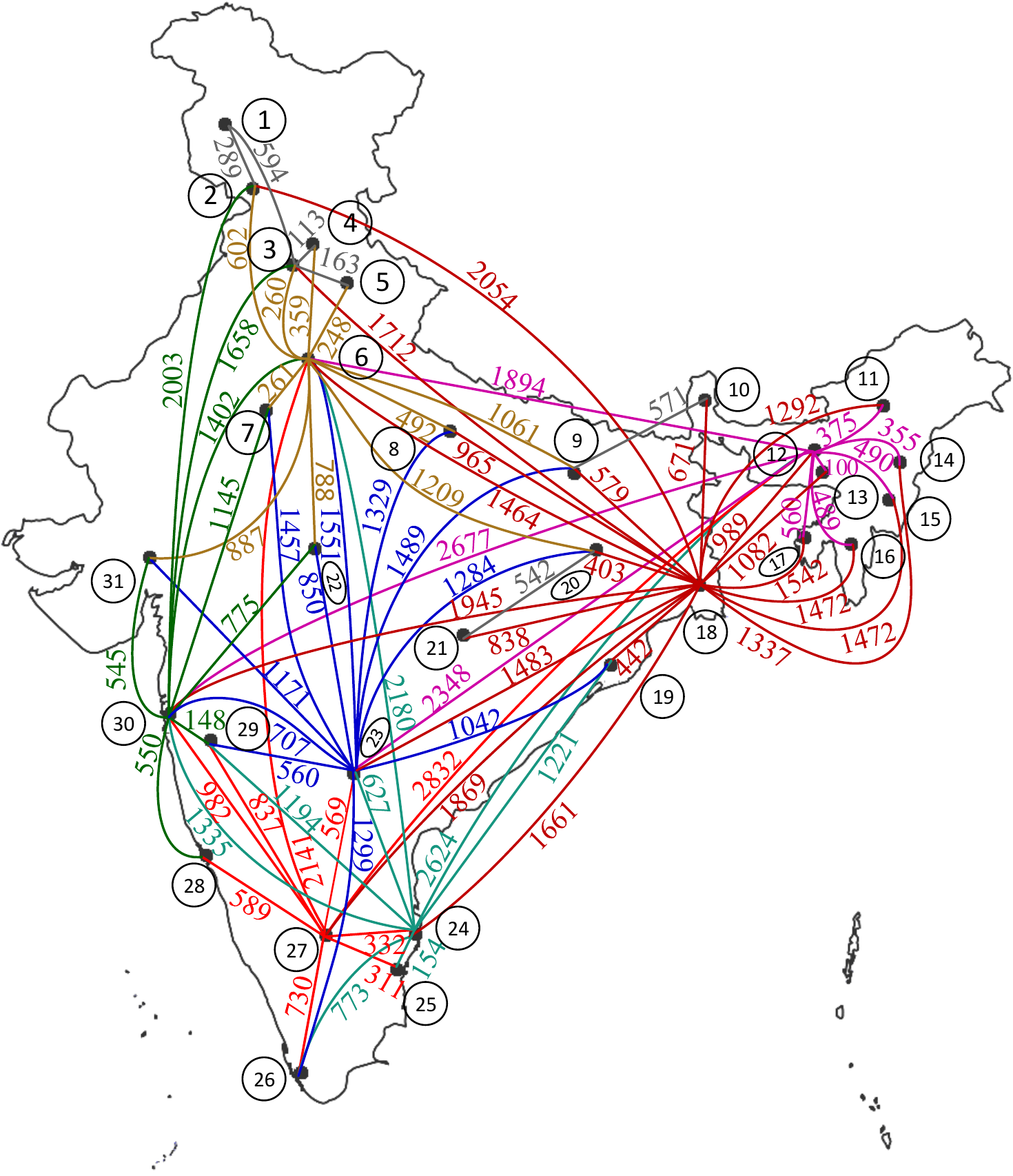}%\hspace*{\fill}%0.53,0.525
% \captionsetup{justification=centering,font={small}}%,margin=1cm,font={footnotesize}, scriptsize, footnotesize
\caption{31-node NKN, India topology. Links are shown with different colors to improve readability~\cite{biswas2019energy}.}
\label{c4:nknPhy}
\end{figure}

To assess the performance of QAG-ENP-IoE heuristic, we consider a realistic large network topology, the \gls{nkn} of India with 31 nodes and 81 fibers (\cref{c4:nknPhy})~\cite{biswas2019energy}. We consider that inline optical amplifiers are placed on fibers 80~km apart, and each IP router port supports 400~Gbps. Each sub-transponder of an SBVT and each VER supports all possible transmission options as presented in \cref{c4:t:lto}. The PC for each IP core router port, SBVT, BV-OXC, VER and amplifier are shown in \cref{c4:t:pcd}. The maximum numbers of SBVTs and VERs that can be installed at any node are considered to be 64 and 3, respectively. Sliceability of an SBVT and the number of SSRs in a VER are fixed at 3 and 16, respectively. It is considered that VERs can be placed only at the top 30\% of the nodes with high degrees. Static traffic demands for all SD node pairs exist, with the exception of the traffic demands within the same nodes (i.e., the diagonal elements of the traffic demand matrix). Traffic demands are considered to be uniformly distributed within [5, 2$X$-5]~Gbps, where $X$ represents the \gls{atd}. The following input parameters: $\alpha$, $\gamma$, $\epsilon$, $\epsilon_{\min}$, $\epsilon_{\max}$, and $\epsilon_{\Delta}$ to be used with QAG-ENP-IoE are given in \cref{c4:t:qalgop}. For each ATD, we execute simulation using QAG-ENP-IoE for ten thousand episodes with a given traffic demand set (matrix), and the best result is taken. For a given value of ATD, we show result averaged over ten different traffic demand sets.

\begin{table}[h!]
\centering
\caption{PC for different network equipment}
\label{c4:t:pcd}
\begin{adjustbox}{max width=\linewidth}%,max height=6cm

\begin{tabular}{@{}rl@{}}
\toprule
 Equipment  &  PC \\ \midrule
 \multirow{1}{*}{IP core router port} & $P_R$ = 560W, $C_R$ = 400~Gbps~\cite{van2012power}\\ \addlinespace%\midrule
 \multirow{1}{*}{SBVT} & As shown in \cref{c4:t:lto}~\cite{biswas2019energy,biswas2019edfa,papanikolaou2015minimizing}\\ \addlinespace
 BV-OXC & 135·$d$+150, $d$ is physical degree of node~\cite{van2012power}\\ \addlinespace
 \multirow{2}{*}{VER} & Lightpath transmit/receive: \cref{c4:t:lto}\\
 & $P_{V_S}$ = 10W, $P_{V_O}$ = 25W~\cite{biswas2019edfa}\\ \addlinespace
 \multirow{1}{*}{Amplifier} & $P_{A_d}$ = 30W, $P_{A_O}$ = 140W~\cite{vizcaino2012energyArticle}\\\bottomrule
\end{tabular}
\end{adjustbox}
\end{table}

\begin{table}[h!]
\centering
\caption{Different parameters for QAG-ENP-IoE}
\label{c4:t:qalgop}
\begin{tabular}{@{}lll@{}}
\toprule
$\alpha$ = 0.1  & $\epsilon$ = 1 & $\epsilon_{min}$ = 0.01 \\
$\gamma$ = 0.99 & $\epsilon_{\Delta}$ = 0.001 & $\epsilon_{max}$ = 1 \\ \bottomrule
\end{tabular}
\end{table}

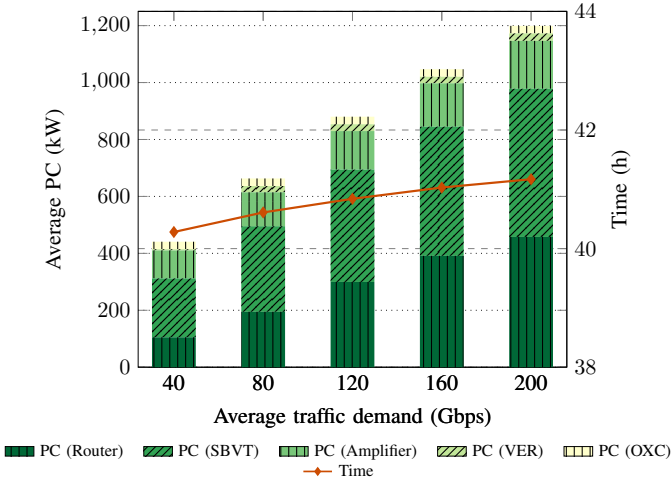
\begin{figure}[h!]%[t]{.57\textwidth}
\centering
\begin{adjustbox}{max width=0.9\linewidth}%{max width=\linewidth,height=\tikzHeight}
	\begin{tikzpicture}
	\pgfplotsset{
		%			scale only axis,
		%			scaled x ticks=base 10:3,
		%			xmin=0, xmax=0.06
		symbolic x coords={40,80,120,160,200},
		xtick=data,
		xlabel=Average traffic demand (Gbps),
		legend columns=5,
		legend style={draw=none},%\scriptsize,,font=\fontsize{6.75}{8}\selectfont,/tikz/every even column/.append style={text width=1.2cm}
		legend to name=Lhgpc,
	}
	
	\begin{axis}[
	%axis y line*=left,
	%		ymin=0, ymax=80,
	ylabel=Average PC (kW),
	ylabel near ticks,
	ybar stacked, 
	ymin=0, ymax=1250,
	ytick={0,200,400,...,1200},
	%		ybar legend,
	bar width=7mm,
	legend style={at={(0.5,-0.2)},anchor=north},
	ymajorgrids=true,
	major grid style={dotted,black},
	]
	
	\addplot [fill=color1,draw=none,area legend,
	postaction={pattern=vertical lines},] 
    coordinates {%{rgb:red,1;green,1;yellow,2.5}
		({40},105.168)
        ({80},194.712)
        ({120},299.600)
        ({160},391.272)
        ({200},457.688)
	};\addlegendentry{PC (Router)}\label{p:hgpc1i}

	\addplot [fill=color2,draw=none,area legend,
	postaction={pattern=north east lines},] coordinates {%{rgb:red,1;green,1;yellow,1}
		({40},206.73504)
        ({80},300.58155)
        ({120},394.45317)
        ({160},453.73572)
        ({200},520.45002)
	};\addlegendentry{PC (SBVT)}\label{p:hgpc1s}
	
	\addplot [fill=color3,draw=none,area legend,
	postaction={pattern=vertical lines},] coordinates {%{rgb:red,1;green,1;yellow,5}
        ({40},98.253)
        ({80},119.145)
        ({120},135.509)
        ({160},152.377)
        ({200},168.292)
	};\addlegendentry{PC (Amplifier)}\label{p:hgpc1a}
	
	\addplot [fill=color4,draw=none,area legend,
	postaction={pattern=north east lines},] coordinates {%{rgb:red,1;green,2.5;yellow,5}
		({40},4.20848)
        ({80},22.10467)
        ({120},23.98063)
        ({160},22.9988)
        ({200},26.18168)
	};\addlegendentry{PC (VER)}\label{p:hgpclv}
	
	\addplot [fill=color5,draw=none,area legend,
	postaction={pattern=vertical lines},] coordinates {%{rgb:red,1;green,2.5;yellow,5}
		({40},26.520)
        ({80},26.520)
        ({120},26.520)
        ({160},26.520)
        ({200},26.520)
	};\addlegendentry{PC (OXC)}\label{p:hgpc1o}
	
	\end{axis}
	
	\begin{axis}[
	set layers,axis background,
	axis y line*=right,
	%		axis x line=none,
	ymin=38, ymax=44,
	ylabel=Time (h),
	ylabel near ticks,
	yticklabel pos=right,
	ymajorgrids=true,
	major grid style={dashed, gray}
	]
	
	%\addlegendimage{empty legend}\addlegendentry{\textbf{Sample:}}
	
	\addlegendimage{/pgfplots/refstyle=p:hgpc1i}\addlegendentry{PC (Router)}
	
	\addlegendimage{/pgfplots/refstyle=p:hgpc1s}\addlegendentry{PC (SBVT)}
	
	\addlegendimage{/pgfplots/refstyle=p:hgpc1a}\addlegendentry{PC (Amplifier)}
	
	\addlegendimage{/pgfplots/refstyle=p:hgpclv}\addlegendentry{PC (VER)}
	
	\addlegendimage{/pgfplots/refstyle=p:hgpc1o}\addlegendentry{PC (OXC)}
	
	\addplot[mark=diamond*,color=color11,line width=1pt]
	coordinates{
		({40},40.28)%1.6942
		({80},40.61)%2.8342
		({120},40.84)%4.3099
		({160},41.03)%4.9639
		({200},41.17)%3.9354
	};\label{p:c4:time}\addlegendentry{Time}
	\end{axis}
	\end{tikzpicture}
\end{adjustbox}
\\
\begin{adjustbox}{max width=\linewidth}%,max height=6cm
	% now generate the legend, using a normal tabular
% \begin{NoHyper}
\begin{tabular}{@{}ccccc@{}}
	\ref{p:hgpc1i} PC (Router) & \ref{p:hgpc1s} PC (SBVT) & \ref{p:hgpc1a} PC (Amplifier) & \ref{p:hgpclv} PC (VER) & \ref{p:hgpc1o} PC (OXC)\\
	\multicolumn{5}{c}{\ref{p:c4:time} Time}
\end{tabular}
% \end{NoHyper}
\end{adjustbox}
\caption{Average PC for different equipment and average execution time for NKN designed with QAG-ENP-IoE.}\label{c4:f:qgpc}
\end{figure}

\cref{c4:f:qgpc} shows the  average overall PC and the PC distribution among different network equipment (SBVT, IP core router, amplifier, VER and OXC) using QAG-ENP-IoE for different ATDs. The overall PC increases with increase in ATD. \cref{c4:f:qgpc} also shows that average execution time increases with increase in ATD. \cref{c4:t:sdpc} shows the standard deviations of PC of different equipment computed for different ATDs. The relative PC (\%) for different equipment are also tabulated in \cref{c4:t:hnsre}. In general, PC due to SBVTs predominates the overall PC of network, followed by PC due to IP core routers, amplifiers, OXCs, and VERs. Relative PC due to amplifiers decreases with increase in ATD, since once all amplifiers on fibers are activated, further increase in PC due to amplifiers cannot take place. Relative PC due to IP core routers increases with increase in ATD, since requirement of IP core router ports increases with increase in ATD.

\begin{table}[h]%{r}{0.5\textwidth}
\captionsetup{labelfont={color=black},font={color=black}}
\caption{Standard deviation in PC of different equipment for NKN}\label{c4:t:sdpc}
\centering
\begin{adjustbox}{max width=\linewidth, }%width=\linewidth,height=1.5cm
\begin{tabular}{@{}lccccc@{}}
\toprule
\textbf{X (Gbps)} & \textbf{40} & \textbf{80} & \textbf{120} & \textbf{160} & \textbf{200} \\ \midrule
Router	    & 1.02	& 3.2	& 6.48	& 12.75	& 10.42	\\
SBVT    	& 1.01	& 6.11	& 6.69	& 12.11	& 13.79	\\
Amplifier	& 0.83	& 6.48	& 5.21	& 5.25	& 6.07	\\
VER	        & 1.22	& 1.80	& 1.45	& 2.04  & 1.36	\\
OXC	        & 0	& 0	& 0	& 0  & 0	\\
Total PC    & 2.79  & 11.03 & 16.81 & 24.28 & 21.38 \\
\bottomrule
\end{tabular}
\end{adjustbox}
\end{table}

\begin{table}[!ht]%{r}{0.5\textwidth}
\caption{Relative PC (\%) for NKN designed with QAG-ENP-IoE}\label{c4:t:hnsre}
\centering
\begin{adjustbox}{}%width=\linewidth
	\begin{tabular}{@{}lccccc@{}}
		\toprule
		\textbf{X (Gbps)} & \textbf{40} & \textbf{80} & \textbf{120} & \textbf{160} & \textbf{200} \\ \midrule
		Router	    & 23.85	& 29.37	& 34.04	& 37.37	& 38.17	\\
        SBVT    	& 46.89	& 45.33	& 44.82	& 43.34	& 43.4	\\
        Amplifier	& 22.29	& 17.97	& 15.4	& 14.56	& 14.03	\\
        OXC	        & 6.02	& 4	    & 3.01	& 2.53	& 2.21	\\
        VER	        & 0.95	& 3.33	& 2.72	& 2.2   & 2.18	\\ \bottomrule
	\end{tabular}
\end{adjustbox}
\end{table}

\cref{f:nlcwwoR} shows the average number of total lightpaths and average number of lightpaths with different capacities set up with different ATDs. With the increase in ATD, more high capacity (e.g., 400 Gbps) lightpaths are required to be set up to efficiently accommodate the increasing traffic demands. Moreover, in general, higher number of lightpaths are set up with increase in ATD. Since PC at SBVT is higher for high capacity lightpaths compared with low capacity lightpaths (\cref{c4:t:lto}), and higher number of lightpaths are set up with increase in ATD, more PC occurs with increase in ATD (as shown in \cref{c4:f:qgpc}). Similarly, with increase in ATD, requirement of different equipment also increases, as reflected in \cref{c4:f:nDE}. As VERs are used only for regeneration and used only at the top 30\% of the nodes with high degrees, the average number of VERs used in the network is very less compared with other equipment.

%\cref{c4:f:nDE} shows the total number of different equipment required in the network. The number of different equipment increases with increase in ATD. 

\newcommand{\barSepSJ}{2.75mm}
\begin{figure}
\centering
\begin{adjustbox}{max width=0.8\linewidth}%,max height=6cm
	\begin{tikzpicture}
	\pgfplotsset{
		symbolic x coords={40,80,120,160,200},
		xtick=data,
		%width=0.75\linewidth
	}
	
	\begin{axis}[
	%axis y line*=left,
	ylabel=Average \# of lightpaths,
	ylabel near ticks,
	ybar stacked,
	ymin=0,ymax=1250,
	ytick={0,200,...,1400},
	bar width=20pt, % reduced bar width
	ymajorgrids=true,
	major grid style={dotted,black},
	xlabel= Average traffic demand (\gbps) % only need one x-label, so move here
	]
	
\pgfplotstableread{
x		y				y-max			y-min
40	    469.5	        23.5	        15.5
80	    206.5	        37.5	        36.5
120	    146.9	        26.1	        25.9
160	    93.1	        19.9	        18.1
200	    83.3	        7.7	            10.3
}{\lcFwV}
	
	\addplot [
	 % added this -- the value depends on the bar shift, bar shift=-\barSepSJ,
	fill=color21, % changed yellow to blue
	draw=none,
	%postaction={pattern=north east lines},
	area legend
	]
	plot [error bars/.cd, y dir=both, y explicit,
	error bar style={line width=0.20pt,solid,black},%,xshift=-\barSepSJ
	error mark options={line width=0.20pt,black,mark size=2.0pt,rotate=90}
	]
	table [y error plus=y-max, y error minus=y-min] {\lcFwV};\label{p:lc40o1} % just a label here

\pgfplotstableread{
x		y				y-max			y-min
40	    474.5	        20.5	        27.5
80	    359.4	        23.6	        19.4
120	    246.7	        27.3	        25.7
160	    194.9	        34.1	        25.9
200	    207.8	        68.2	        55.8
}{\lcHwV}
	
	% added negative bar shift here as well
	\addplot [fill=color22,draw=none,
	%postaction={pattern=north east lines},bar shift=-\barSepSJ,
	area legend] 
	plot [error bars/.cd, y dir=both, y explicit,
	error bar style={line width=0.20pt,solid,black,},%xshift=-\barSepSJ
	error mark options={line width=0.20pt,black,mark size=2.0pt,rotate=90}
	]
	table [y error plus=y-max, y error minus=y-min] {\lcHwV};\label{p:lc100o1} % just a label here
	
\pgfplotstableread{
x		y				y-max			y-min
40	    31	            2	            20
80	    444.3	        31.7	        49.3
120	    516.2	        23.8	        43.2
160	    356.9	        28.1	        37.9
200	    294.2	        25.8	        28.2
}{\lcTwV}
	
	% added negative bar shift here as well
	\addplot [fill=color28,draw=none,
	%postaction={pattern=north east lines},bar shift=-\barSepSJ,
	area legend] 
	plot [error bars/.cd, y dir=both, y explicit,
	error bar style={line width=0.20pt,solid,black},%,xshift=-\barSepSJ
	error mark options={line width=0.20pt,black,mark size=2.0pt,rotate=90}
	]
	table [y error plus=y-max, y error minus=y-min] {\lcTwV};\label{p:lc200o1} % just a label here

\pgfplotstableread{
x		y				y-max			y-min
40	    9	            4	            7
80	    12	            1	            8
120	    181.1	        19.9	        28.1
160	    451	            20	            37
200	    580.7	        28.3	        37.7
}{\lcFhwV}
	
	% added negative bar shift here as well
	\addplot [fill=color32,draw=none,
	%postaction={pattern=north east lines},bar shift=-\barSepSJ,
	area legend] 
	plot [error bars/.cd, y dir=both, y explicit,
	error bar style={line width=0.20pt,solid,black},%,xshift=-\barSepSJ
	error mark options={line width=0.20pt,black,mark size=2.0pt,rotate=90}
	]
	table [y error plus=y-max, y error minus=y-min] {\lcFhwV};\label{p:lc400o1} % just a label here
	\end{axis}
	\end{tikzpicture}
\end{adjustbox}

\begin{adjustbox}{max width=\linewidth}%,max height=6cm
	% now generate the legend, using a normal tabular
% \begin{NoHyper}
	\begin{tabular}{@{}ccccc@{}}
		Lightpath capacity (\gbps): & \ref{p:lc40o1} 40 & \ref{p:lc100o1} 100 & \ref{p:lc200o1} 200 & \ref{p:lc400o1} 400
	\end{tabular}
% \end{NoHyper}
\end{adjustbox}
% \captionsetup{labelfont={color=black}}
\caption{Average number of total lightpaths and average number of lightpaths with different capacities set up for NKN designed with QAG-ENP-IoE.}
\label{f:nlcwwoR}
\end{figure}

\begin{figure}[h!]
	\centering
	\begin{adjustbox}{max width=0.8\linewidth}%,max height=6cm
		\begin{tikzpicture}
		\pgfplotsset{
			symbolic x coords={40,80,120,160,200},
			xtick=data,
			%width=0.75\linewidth
		}
		
		\begin{axis}[
		%axis y line*=left,
		ylabel=Average \# of equipment,
		ylabel near ticks,
		ybar stacked,
		bar width=20pt,
		ymin=0,ymax=2500,
		%bar width=\barWidthSJ, % reduced bar width
		ymajorgrids=true,
		major grid style={dotted,black},
		xlabel= Average traffic demand (Gbps), 
		xlabel style={align=center,black},
		ylabel style={align=center,black},% only need one x-label, so move here
		]
		
\pgfplotstableread{
x		y				y-max			y-min
40  	187.8	        3.19	        2.8
80  	347.7	        10.3	        9.7
120 	535	            19	            16
160 	698.7	        27.3	        30.7
200	    817.3	        34.7	        29.3
}{\nrE}
		
% added negative bar shift here as well
\addplot [
%bar shift=-\barSepSJ, % added this -- the value depends on the bar shift
fill=color16, % changed yellow to blue
draw=none,
%postaction={pattern=north east lines},
area legend
] 
plot [error bars/.cd, y dir=both, y explicit,
error bar style={line width=0.20pt,solid,black},
error mark options={line width=0.20pt,black,mark size=2.0pt,rotate=90}
]
table [y error plus=y-max, y error minus=y-min] {\nrE};\label{p:nrE} % just a label here
%xshift=-\barSepSJ

\pgfplotstableread{
x		y				y-max			y-min
40	    310.5	        1.5	            0.5
80	    386.3	        13.7	            34.3
120	    539.8	        16.2	            13.8
160	    681.2	        30.8	            30.2
200	    800.6	        37.4	            39.6
}{\nstE}

% added negative bar shift here as well
\addplot [%bar shift=-\barSepSJ,
fill=color17,draw=none,
%postaction={pattern=north east lines},
area legend] 
plot [error bars/.cd, y dir=both, y explicit,
error bar style={line width=0.20pt,solid,black,},
error mark options={line width=0.20pt,black,mark size=2.0pt,rotate=90}
]
table [y error plus=y-max, y error minus=y-min] {\nstE};\label{p:nstE} % just a label here
%xshift=-\barSepSJ

\pgfplotstableread{
x		y				y-max			y-min
40	    310.1	        0.9	            0.1
80	    386.6	        13.4	            28.6
120	    539.3	        13.7	            14.3
160	    680.6	        27.4	            30.6
200	    800.4	        33.6	            40.4
}{\nsrE}

% added negative bar shift here as well
\addplot [%bar shift=-\barSepSJ,
fill=color18,draw=none,
%postaction={pattern=north east lines},
area legend] 
plot [error bars/.cd, y dir=both, y explicit,
error bar style={line width=0.20pt,solid,black,},
error mark options={line width=0.20pt,black,mark size=2.0pt,rotate=90}
]
table [y error plus=y-max, y error minus=y-min] {\nsrE};\label{p:nsrE} % just a label here
%xshift=-\barSepSJ

\pgfplotstableread{
x		y				y-max			y-min
40	    5.6	            2.4	            3.6
80	    18.3        	2.7	            2.3
120	    21.9	        2.1	            1.9
160	    24.1        	1.9	            2.1
200	    25.5	        2.5	            1.5
}{\nvE}

% added negative bar shift here as well
\addplot [%bar shift=-\barSepSJ,
fill=color19,draw=none,
%postaction={pattern=north east lines},
area legend] 
plot [error bars/.cd, y dir=both, y explicit,
error bar style={line width=0.20pt,solid,black,},
error mark options={line width=0.20pt,black,mark size=2.0pt,rotate=90}
]
table [y error plus=y-max, y error minus=y-min] {\nvE};\label{p:nvE} % just a label here
%xshift=-\barSepSJ

		\end{axis}
		\end{tikzpicture}
	\end{adjustbox}
	
	\begin{adjustbox}{max width=\linewidth}%,max height=6cm
		% now generate the legend, using a normal tabular
% 	\begin{NoHyper}
		\begin{tabular}{@{}cccc@{}}
			\ref{p:nrE} IP core router port & \ref{p:nstE} SBVT (Tx) & \ref{p:nsrE} SBVT (Rx) & \ref{p:nvE} VER
		\end{tabular}
% 	\end{NoHyper}
	\end{adjustbox}
% 	\captionsetup{labelfont={color=black},font={color=black}}
	\caption{Average number of different equipment installed for NKN designed with QAG-ENP-IoE.}
	\label{c4:f:nDE}
\end{figure}

The performance of QAG-ENP-IoE is compared with the following different heuristic methods:

\renewcommand\labelenumi{(\roman{enumi})}
\begin{enumerate}
    \item Shortest path network planning with traffic provisioning in descending order (SP).
    \item Network planning using mAG-ENP-IoE with traffic provisioning in descending order (D-GH).
    \item Network planning using mAG-ENP-IoE with traffic provisioning in ascending order (A-GH).
    %\item Network planning using m-AGENP-IoE with bandwidth provisioning from node with the lowest index to highest index \textcolor{red}{(I-GH)}.
    \item Network planning using mAG-ENP-IoE with traffic provisioning following node indices [starting from the first element to the last element of traffic demand matrix] (I-GH).
\end{enumerate}

\renewcommand{\barSepSJ}{1mm}
\begin{figure}[!b]
\centering
\begin{adjustbox}{max width=1\linewidth}%{max width=\linewidth,height=\tikzHeight}
\begin{tikzpicture}
\pgfplotsset{
    height=5cm,%\tikzpH
	width=\linewidth,%\tikzpW,
	symbolic x coords={40,80,120,160,200},
	xtick=data,%{40,80,120,160,200},%data,
	xlabel=Average traffic demand (Gbps),
	legend columns=5,
	legend style={draw=none},
	legend to name=Lmhgc,
}

\begin{axis}[
%axis y line*=left,
ylabel=Average PC (kW),
ylabel near ticks,
ybar,
try min ticks = 7,
ymin=400,ymax=1250,
ytick={400,600,...,1200},
bar width=1mm,
legend style={at={(0.5,-0.2)},anchor=north},
grid=both,%hide x axis,
major grid style={dotted,black},
]

\pgfplotstableread{
x		y				y-max			y-min
40	    540.8845	    13.5977	        14.2403
80	    756.0632	    12.4622	        18.0614
120	    920.0628	    23.4742	        31.0732
160	    1096.9035	    28.7595	        38.5127
200	    1213.1317	    19.1909	        20.8285
}{\mytable}
\addplot [%bar shift=-\barSepSJ, % added this -- the value depends on the bar shift
fill=cfp6o1, % changed yellow to blue
draw=none,
postaction={pattern=north east lines},
	%%		error bars/.cd, y dir=both, y explicit,
area legend] 
plot [error bars/.cd, y dir=both, y explicit, error bar style={line width=0.20pt,solid,black,}, error mark options={line width = 0.20pt, black, mark size = 2.0pt, rotate=90}]%error bar style={xshift=-\barSepSJ},
table [y error plus=y-max, y error minus=y-min] {\mytable};\label{c4:p:sp}

\pgfplotstableread{
x		y				y-max		y-min
40	    470.3037	    13.9857	    12.5023
80	    670.0793	    32.8873	    21.1653
120	    884.4086	    15.1084	    10.5266
160	    1018.3908	    0	        0
200	    0	            0	        0
}{\cfourpmag}
\addplot [%bar shift=\barSepSJ,
postaction={pattern=north west lines},
fill=cfp6o2,draw=none,area legend]
plot [error bars/.cd, y dir=both, y explicit, error bar style={line width=0.20pt,solid,black,}, error mark options={line width = 0.20pt, black, mark size = 2.0pt, rotate=90}]
table [y error plus=y-max, y error minus=y-min]{\cfourpmag};\label{c4:p:mag}

\pgfplotstableread{
x		y				y-max		y-min
40	    470.7341	    17.5905	    15.9741
80	    693.3997	    21.3699	    13.8573
120	    922.1323	    28.7425	    22.5971
160	    1085.4562	    0	        0
200	    0	            0	        0
}{\cfourpqag}
\addplot [%bar shift=\barSepSJ,
postaction={pattern=north east lines},
fill=cfp6o3,draw=none,area legend]
plot [error bars/.cd, y dir=both, y explicit, error bar style={line width=0.20pt,solid,black,}, error mark options={line width = 0.20pt, black, mark size = 2.0pt, rotate=90}]
table [y error plus=y-max, y error minus=y-min]{\cfourpqag};\label{c4:p:qag}

\pgfplotstableread{
x		y				y-max		y-min
40	    482.6632	    8.9202	    5.8232
80	    705.9999	    9.2441	    17.0587
120	    885.8728	    0	        0
160	    1065.2166	    0	        0
200	    0	            0	        0
}{\cfourpter}
\addplot [%bar shift=\barSepSJ,
postaction={pattern=north west lines},
fill=cfp6o4,draw=none,area legend]
plot [error bars/.cd, y dir=both, y explicit, error bar style={line width=0.20pt,solid,black,}, error mark options={line width = 0.20pt, black, mark size = 2.0pt, rotate=90}]
table [y error plus=y-max, y error minus=y-min]{\cfourpter};\label{c4:p:ter}

\pgfplotstableread{
x		y				y-max		y-min
40	    452.6477	    3.0875	    4.4843
80	    663.9555	    16.7899	    16.2223
120	    882.9102	    29.5191	    21.2751
160	    1061.6521	    42.6637	    25.1779
200	    1197.1837       43.9681	    40.8775
}{\cfourga}
% \addplot [%bar shift=\barSepSJ,
% postaction={pattern=north east lines},
% fill=cfp6o5,draw=none,area legend]
% plot [error bars/.cd, y dir=both, y explicit, error bar style={line width=0.20pt,solid,black,}, error mark options={line width = 0.20pt, black, mark size = 2.0pt, rotate=90}]
% table [y error plus=y-max, y error minus=y-min]{\cfourga};\label{c4:p:ga}

%%%%%%%%%%%%%%
\pgfplotstableread{
x		y				y-max		y-min
40	    440.8845	    3.5977	    4.2403
80	    663.0632	    12.4622	    18.0614
120	    880.0628	    23.4742	    21.0732
160	    1046.9035	    38.7595	    28.5127
200	    1199.1317	    29.1909	    40.8285
}{\cfourpter}
\addplot [%bar shift=\barSepSJ,
postaction={pattern=north west lines},
fill=cfp6o6,draw=none,area legend]
plot [error bars/.cd, y dir=both, y explicit, error bar style={line width=0.20pt,solid,black,}, error mark options={line width = 0.20pt, black, mark size = 2.0pt, rotate=90}]
table [y error plus=y-max, y error minus=y-min]{\cfourpter};\label{c4:p:best}

\end{axis}

\end{tikzpicture}
\end{adjustbox}
\\
\begin{adjustbox}{max width=.75\linewidth}
%\ref{Lmhgc}
% \begin{NoHyper}
\begin{tabular}{@{}rccccc@{}}
	& SP & D-GH & A-GH & I-GH & QAG-ENP-IoE\\
	PC: & \ref{c4:p:sp} & \ref{c4:p:mag} & \ref{c4:p:qag} & \ref{c4:p:ter} & \ref{c4:p:best}\\%$\perp$
%	Time: & \ref{c4:t:sp} & \ref{c4:t:mag} & \ref{c4:t:qag}& \ref{c4:t:ter}& \ref{c4:t:best}
\end{tabular}
% \end{NoHyper}
\end{adjustbox}
\captionof{figure}{Average PC for network designed with SP, D-GH, A-GH, I-GH and QAG-ENP-IoE.}.\label{c4:f:mhgc}
\end{figure}

In \cref{c4:f:mhgc}, we show the average PC for network designed with SP, D-GH, A-GH, I-GH, and QAG-ENP-IoE. SP heuristic first explores the existing lightpaths with free capacity for provisioning the selected traffic. In case, such lightpaths are not available, a new lightpath is set up following the shortest distance path between SD node pair of the traffic, and the traffic is provisioned through it. In all cases (i.e., all ATD values), PC for network designed using SP is the highest as the method does not consider PC during network designing. With 40-120 Gbps ATD, D-GH, A-GH and I-GH offer almost 4-15\%, 0-15\% and 5-12\% improvement in average PC compared with SP, respectively. Network designing is not feasible using D-GH, A-GH and I-GH for none of the considered traffic demand matrices with ATD of 200 Gbps, since resources are exhausted at one or more locations. These three methods can provide solution for only one traffic demand matrix with ATD of 160 Gbps, due to exhaustion of resources. QAG-ENP-IoE learns the best sequence for energy-efficient network planning. Even though the heuristic may fail to provide feasible solution for a given traffic demand matrix in some cases (out of the given ten thousand episodes) due to paucity of resources, the heuristic eventually can always successfully provision the entire traffic demand matrix. The heuristic comes up with an optimal sequence, i.e., a policy that provisions traffic for energy-efficient IP-over-EON. QAG-ENP-IoE offers 1-23\%, 5-7\%, 1-7\% and 1-9\% improvement in PC values compared with SP, D-GH, A-GH and I-GH methods, respectively. In \cref{c3:f:sr}, the average number of feasible solutions, i.e., the average number of success per 1000 episodes are plotted. Initially, the number of success is low, and in general, the number of success increases with increase in the index of 1000 episodes.

{

\renewcommand{\barSepSJ}{2.75mm}
\begin{figure}[!h]
\centering
\begin{adjustbox}{max width=\linewidth}%,max height=6cm
	\begin{tikzpicture}
	\pgfplotsset{
		height=5cm,%\tikzpH
		width=\linewidth,%\tikzpW,
		%symbolic x coords={40,80,120,160,200},
		xtick=data,
		%width=0.75\linewidth
	}
	
	\begin{axis}[
	%axis y line*=left,
	ylabel near ticks,
	ylabel={Average \# of success\\ per 1000 episodes},
	ylabel style={align=center,black},
	ybar stacked,
	ymin=550,ymax=950,
	ytick={550,650,...,950},
	bar width=15pt, % reduced bar width
	ymajorgrids=true,
	major grid style={dotted,black},
	xlabel= Index of 1000 episodes% only need one x-label, so move here
	]
	
\pgfplotstableread{
x		y			y-max			y-min
1	    637         92	            36
2	    688.2       107.8	        23.6
3	    732.4       124	            28.4
4	    694.5       146.5	        24.5
5	    800	        81	            104
6	    844	        46	            087
7	    870	        27	            153
8	    883	        15	            136
9	    878	        21	            139
10	    902	        29	            134
}{\lcFwV}
	
	\addplot [
	 % added this -- the value depends on the bar shift, bar shift=-\barSepSJ,
	fill=cd44, % changed yellow to blue
	draw=none,
	%postaction={pattern=north east lines},
	area legend
	]
	plot [error bars/.cd, y dir=both, y explicit,
	error bar style={line width=0.20pt,solid,black},%,xshift=-\barSepSJ
	error mark options={line width=0.20pt,black,mark size=2.0pt,rotate=90}
	]
	table [y error plus=y-max, y error minus=y-min] {\lcFwV};\label{p:lc40o1} % just a label here
	\end{axis}
	\end{tikzpicture}
\end{adjustbox}
% \captionsetup{labelfont={color=black}}
\caption{Average number of success per 1000 episodes.}
\label{c3:f:sr}
\end{figure}
}

%(i.e., \textcolor{red}{i.e., $n$} in the $n^{th}$ 1000 episode)

% Out of these three strategies, s-AG is most affected due to exhaust of resources and overall PC is also higher in case of successful completion.
% lh-AG strategy gives lowest overall PC and highest number of successful completion among the three. 

%\cref{f:avgpcw}

% \begin{figure}
% \centering
% \begin{adjustbox}{max width=\linewidth}%,max height=6cm
% \renewcommand{\barSepSJ}{2.75mm}
% \input{./Fig/avgPCwindow1000}
% \end{adjustbox}
% \captionsetup{labelfont={color=black}}
% \caption{Average PC per 1000 episodes.}
% \label{f:avgpcw}
% \end{figure}

\section{Conclusion}\label{c4:conclusion}
In this paper, network planning implemented through efficient resource and static traffic demand provisioning for energy-efficient VER-assisted IP-over-EON is explored. We propose RL (in particular, Q-learning) and AG-based heuristic for energy-efficient resource allocation and traffic provisioning. Even though the heuristic may fail to provide feasible solution for a given traffic demand matrix in some cases (out of the given ten thousand episodes) due to paucity of resources, the heuristic eventually can always successfully provision the entire traffic demand matrix. Simulation results show the increase in PC with the increase in ATD (i.e., the traffic volume). The PC in SBVTs and IP core routers is found to predominate over the PC in other equipment. The proposed heuristic offers up to 23\% and up to 9\% improvement in PC compared with SP and AG-based greedy heuristics (viz., D-GH, A-GH and I-GH) respectively.

%Result shows that the use of VER along with using D-GH can reduce PC of network up to 7.33\% \textcolor{red}{compared to the case when VER is not used}.

%The AG-based energy-efficient greedy method is used to accommodate a given traffic element, so as to accommodate the traffic with the least increase in PC. The Q-learning method is used to find the suitable sequence of traffic \textcolor{red}{provision} such that the overall PC in the network reduces.
\balance
%In this paper, optimization of these initial parameters are not considered.
%\bibliographystyle{IEEEtran}
%\bibliography{EON\string_Master}
\printbibliography[title=References]

\balance
% \DeclareBibliographyCategory{Online}
% \bibliography{EON\string_Master}

\cmmnt{
\section*{Author Biographies}
\setlength\intextsep{0pt}

\begin{wrapfigure}{L}{0.21\textwidth}
\includegraphics[width=0.20\textwidth]{AuthBio/biswas2020.jpg}
\end{wrapfigure}
\paragraph{}
\noindent \textbf{Pramit Biswas} received his B.Tech and M.Tech degree in electrical engineering from Kalyani Government Engineering College, West Bengal, India, in 2012 and University College of Science and Technology (Rajabazar), University of Calcutta, West Bengal, India, in 2014 respectively. 

He is currently working toward his PhD degree in the Department of Electrical Engineering at the Indian Institute of Technology Patna, Bihar, India. He also carried part of his research work from G.S Sanyal School of Telecommunication at the Indian Institute of Technology Kharagpur, West Bengal, India. His present research interests include but not limited to the fields of cost and energy-efficient optical network design and operation, wireless-optical access networks and software-defined networks.

% Mr. Biswas received Visvesvaraya Fellowship for his doctoral research.

\begin{wrapfigure}{L}{0.21\textwidth}
\includegraphics[width=0.20\textwidth]{AuthBio/Akhtar2021.jpg}
\end{wrapfigure}
\paragraph{}
\noindent \textbf{Md Shahbaz Akhtar} received his B.Tech degree in Electronics and Communication Engineering from GGSIPU University, Delhi, India, in 2013, and his M.Tech degree from IIT Dhanbad, Jharkhand, India, in 2015. He is currently working toward his Ph.D. degree in the Department of Electrical Engineering at the Indian Institute of Technology Patna, Patna, Bihar, India. His research interests are in the fields of computer communication and networks, mobile communications, survivable and energy-efficient next-generation passive optical access networks, wireless cellular networks and hybrid wireless-optical access networks.

\begin{wrapfigure}{L}{0.21\textwidth}
\includegraphics[width=0.20\textwidth]{AuthBio/Adhya2018.jpg}
\end{wrapfigure}
\paragraph{}
\noindent \textbf{Aneek Adhya} received his B.Tech degree in Electronics and Communication Engineering from Kalyani Government Engineering College, Kalyani, India, in 2001, and his M.Tech degree from the University College of Science and Technology, University of Calcutta, Calcutta, India, in 2003. He received his Ph.D. from the Electronics and Electrical Communication Engineering Department of the Indian Institute of Technology Kharagpur, Kharagpur, India, in 2009. 

He served as an Assistant Professor in the Electrical Engineering Department of the Indian Institute of Technology Patna, India from 2009 to 2017. Since May 2017, he has been serving as an Assistant Professor in the G. S. Sanyal School of Telecommunications, Indian Institute of Technology Kharagpur, India. He has authored or co-authored many international peer-reviewed journals, conferences and technical papers. He also serves as reviewer of several renowned peer-reviewed journals and conferences in the broad area of optical communications and optical networking. His research interests include computer communication and networks, elastic optical networks, energy efficiency in backbone and access networks, physical layer impairment-aware network design and hybrid wireless-optical broadband access networks.

\begin{wrapfigure}{L}{0.21\textwidth}
\includegraphics[width=0.20\textwidth]{AuthBio/majhi2019.jpg}
\end{wrapfigure}
\paragraph{}
\noindent \textbf{Sudhan Majhi} received the M.Tech. degree in computer science and data processing from IIT Kharagpur, Kharagpur, India, in 2004, and the Ph.D. degree from Nanyang Technological University, Singapore, in 2008.

He is currently an Associate Professor with the Department of Electrical Engineering, IIT Patna, Patna, India, where he is a Fellow of Sir Visvesvaraya Young Faculty Research. He has a postdoctoral experience with the University of Michigan, Dearborn, MI, USA; Institute of Electronics and Telecommunications, Rennes, France; and Nanyang Technological University, Singapore. His current research interests include signal processing for wireless communication which includes blind synchronization and parameter estimation, cooperative communications, physical layer security for cognitive radio, sequence design, orthogonal frequency-division multiplexing (OFDM), multiple-input multiple-output (MIMO), single carrier-frequency division multiple access, and MIMO OFDM.

% \balance

\begin{wrapfigure}{L}{0.21\textwidth}
\includegraphics[width=0.20\textwidth]{AuthBio/saha2020.jpg}
\end{wrapfigure}
\paragraph{}
\noindent \textbf{Sriparna Saha} (Senior Member, IEEE) received the master’s and Ph.D. degrees in computer science from the Indian Statistical Institute, Kolkata, India, in 2005 and 2011, respectively.

She is currently an Associate Professor with the Department of Computer Science and Engineering, IIT Patna, Patna, India. She has authored or coauthored more than 120 articles. Her current research interests include machine learning, pattern recognition, multiobjective optimization, language processing, and biomedical information extraction.

Dr. Saha was a recipient of several awards, including the Lt Rashi Roy Memorial Gold Medal from the Indian Statistical Institute for outstanding performance in M.Tech. (computer science), the Google India Women in Engineering Award in 2008, the NASI Young Scientist Platinum Jubilee Award in 2016, the BIRD Award in 2016, the IEI Young Engineer’s Award in 2016, the SERB Women in Excellence Award in 2018, the SERB Early Career Research Award in 2018, the Humboldt Research Fellowship, the Indo–U.S. Fellowship for Women in STEMM (WISTEMM) Women Overseas Fellowship Program in 2018, and the CNRS Fellowship. She was a recipient of the India4EU Fellowship of the European Union to work as a Post-Doctoral Research Fellow at the University of Trento, Trento, Italy, from September 2010 to January 2011, and the Erasmus Mundus Mobility with Asia (EMMA) Fellowship of the European Union to work as a Post-Doctoral Research Fellow at Heidelberg University, Heidelberg, Germany, from September 2009 to June 2010.
}

\end{document}